# Coupled Plasmon Modes in 2D Gold Nanoparticle Clusters for Local Temperature Control

Rituraj Borah[1], Sammy W. Verbruggen[1,*]

[1]Sustainable Energy, Air & Water Technology (DuEL), Department of Bioscience Engineering,

University of Antwerp, Groenenborgerlaan 171, 2020 Antwerp, Belgium

[*]Sammy.Verbruggen@uantwerpen.be

**Abstract**

Assembly of nanostructures exhibit collective plasmonic response due to coupling of the individual plasmon modes of the constituent nanostructures. In the context of self-assembly of nanoparticles, closed packed 2D cluster of spherical nanoparticles is an important composite system that promises numerous applications. Thus, the present study probes the collective plasmonic characteristics and resulting photothermal effects in closed packed 2D Au nanoparticle clusters in order to delineate the effect of cluster size, inter-particle distance and particle size. Smaller nanoparticles (20 nm and 40 nm in diameter) that exhibit low individual scattering and high absorption were considered for their relevance to photothermal applications. Based on rigorous validation of the numerical model, the present study compares the optical response of clusters of different sizes from 1 nanoparticle up to 61 nanoparticles. In addition to the change of spectral position and characteristics with increasing cluster size, an optimal cluster size for maximum absorption per nanoparticle is also determined for enhanced photothermal effects. The effect of particle size and inter-particle distance is investigated in order to elucidate the nature of interaction in terms of near-field and far-field coupling. The photothermal effects resulting from absorption is compared for different cluster sizes and inter-particle distances considering a homogeneous water medium. A strong dependence of the steady state temperature of the nanoparticles on the cluster size, particle position in the cluster, incident light polarization and inter-particle distance promises interesting local temperature control applications.

## 1. Introduction

Localized surface plasmon resonance (LSPR) in metal nanoparticles under the influence of an oscillating electric field results in strong enhancement of this field in the direct vicinity of the nanoparticle and hence, tight optical confinement beyond the diffraction limits.[1] Be it concentration of light, spectral characteristics or thermal dissipation, various aspects of plasmonics have gained significant scientific attention for their application in different areas such as photocatalysis,[2-4] sensing,[5] solar energy devices,[6] imaging,[7] etc.[8,9] [10]Apart from the optical properties of the material, since the plasmonic behavior of a nanostructure is also strongly influenced by the size and geometrical aspects, design and development of novel plasmonic nanostructures is a primary theme of current plasmonic research.[11-13]

One structural aspect that has gained a lot of attention is the collective plasmonic behavior of nanoparticle/nanostructure assemblies.[14,15] Interaction among the plasmon modes of individual components of a composite nanostructure gives rise to interesting optical characteristics such as higher order modes, magnetic resonance[16] and fano resonance[17] that are promising for various applications.[18,19] For instance, fano resonance can be useful in chemical and biological sensing, electro-optics, switching, etc.[20] Thus, considerable efforts have been devoted towards the understanding and fabrication of such composite nanostructures.[21-24] Obviously, the discussion on collective plasmonic behavior starts with simple dimers for which many past studies have shed light on relevant aspects such as the polarization dependence, near-field enhancement, spectral shift, and the effect of inter-particle distance.[25-28] Apart from the classical electromagnetic field theory, the analogy of molecular hybridization theories with the interaction of plasmon modes has yielded an alternative approach to explain the collective plasmonic behavior of composite nanostructures.[29,30] Importantly, the spectral red shift and the significant enhancement of the near-field between the particles are striking features of dimers. As one adds more nanoparticles to a linear arrangement, the red shift gets stronger with increasing chain length.[31,32] On the other hand, it is interesting to note that an equilateral 2D arrangement of three nanoparticles exhibits little dependence on the in-plane polarization angle of incident light. This planer symmetry is even further extended to 3D isotropy for 3D symmetric tetramers. One important aspect of many particle systems is the interaction among the radiative modes that results in constructive and destructive interference often indicated by the Fano-like characteristic dip in the scattering spectra. In the discussion of nanoparticle clusters, an obvious question arises regarding the infinity limit beyond which the cluster size becomes insignificant. The formalism given by Modinos *et al*. for the estimation of optical response of infinite arrays of nanoparticles is based on the methods developed in relation to electron scattering by two-dimensional atomic layers.[33-35] The estimated optical spectra from this formulation has been shown to agree reasonable well with the experimental spectra for gold nanoparticle lattices. An alternative approach among others is the Korringa–Kohn–Rostoker (KKR) method which has been demonstrated as a rather robust approach for electromagnetic waves in a periodic dielectric medium in their full complexity.[36,37] Zundel and Manjavacas discussed this infinity limit for large 100 nm nanoparticles arranged in a square array, a system with strong far-field coupling.[38] It is intuitively expected that this limit should be dependent on the nanostructure size, geometry and arrangement.

For the fabrication of nanoparticle/nanostructure oligomers, techniques like e-beam lithography or ion beam milling are widely used due to their high controllability.[39,40] However, these techniques are expensive and inefficient at small dimensions. On the other hand, formation of nanoparticle clusters with precise dimensions by self-assembly techniques is a challenging task and an important area of investigation.[41,44] The thermo-plasmonic effect and resulting phenomena such as plasmo-fluidic convection, phase transition, vapor generation, etc., in nanoparticles and nanoparticle arrays have gained considerable attention over the years.[45-53] In view of the immense possibilities and rapid advances in the area of nanostructure design and applications, promising research avenues are continuously emerging.

Pertaining to self-assembled structures, plasmonic response of clusters of small (<50 nm) nanoparticles is an important unexplored aspect of investigation as such systems promise many applications. Particularly due to low scattering and high absorption, small nanoparticle systems are very relevant to photothermal applications. Based on rigorous validation of the numerical methods used, this study sheds light on collective plasmonic behavior and resulting photothermal effects in 2D closed packed nanoparticle clusters of small (20 and 40 nm in diameter) Au nanoparticles. Clusters of varying size and inter-particle distance were considered in order to delineate the effect of these structural aspects on the collective plasmonic behavior. Apart from the optical response, the steady state temperatures of the nanoparticles in water medium were calculated in order to evaluate the photothermal characteristics of these clusters. Our simulations show a strong dependence of the collective resonance on the structural parameters of the cluster. This results in an interesting optical response and associated photothermal effects. These insights will be valuable in the development of plasmonic nanostructures for a myriad of applications.

## 2. Problem specification and numerical methodology

The present study involves electromagnetic simulations of closed packed 2D assemblies of Au nanoparticles with varying number of nanoparticles and inter-particle distance (Figure 1). A Finite Element Modeling (FEM) based numerical framework in COMSOL Multiphysics was used to numerically solve the frequency domain form of Maxwell's equations (eq. 1):

$$\nabla \times (\mu_r^{-1} \nabla \times E_{sc}) - k_o^2 (\varepsilon_r - j\frac{\sigma}{\omega \varepsilon_o}) E_{sc} = 0 \qquad \text{(eq. 1)}$$

Where $\mu_r, \varepsilon_r,$ and $\sigma$ are material properties namely relative permeability, relative permittivity and electrical conductivity respectively. Eq. 1 is solved numerically in the entire computational domain for the scattered electric field, $E_{sc}$. The scattered electric/magnetic fields, $E_{sc}/H_{sc}$ superposed with the incident electric/magnetic fields, $E_{inc}/H_{inc}$ forms the total field $E/H$. A spherical computational domain with a perfectly matched layer was implemented. The optical constants for Au were taken from Johnson and Christy.[54] An adjacent substrate with a higher refractive index induces a relatively weak spectral shift. This effect is stronger for greater contact between the nanoparticle and the substrate.[55,56] Although the effect of the substrate can be an important aspect to consider, its inclusion in the model entails greater consumption of computational resources since the background electric field in absence of the nanostructure has to be first evaluated before solving for the scattered field in presence of the nanostructure.[57] In view of the complexity of the structures considered herein, the substrate effect is excluded in order to optimize the computational resources while limiting the validity of the present results to low reflecting substrates. Thus, the problem simplifies to solving for the scattered field in a homogeneous medium in the presence of a plane polarized incident electric field. The minimum inter-particle distance considered in this study is 1 nm for which quantum tunneling effects can be neglected.[58] The computational domain was discretized with tetrahedral elements in the physical domain *i.e.*, the nanoparticles and the surrounding medium. While the perfectly matched layer was discretized by prismatic elements.

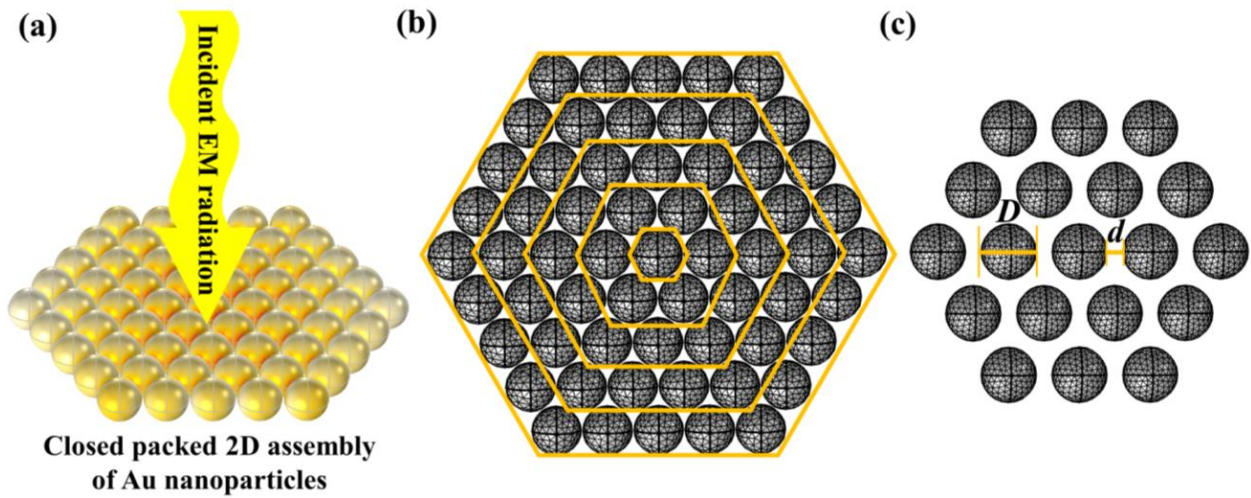

Figure 1. Schematic representation of (a) nanoparticle 2D nanoparticle cluster irradiated with plane polarizd electromagnetic wave, (b) top view of nanoparticle cluster with hexagonal yellow boundaries defining different cluster sizes considered. The yellow boundaries starting from the outermost one define clusters of 61, 37, 19, 9 and single nanoparticle respectively. (c) A cluster of 19 nanoparticles with the definition of inter-particle distance *d*.

The optical cross sections were obtained from the scattered field solution. Absorption and scattering cross sections for a nanoparticle can be defined as:

$$\sigma_{abs} = \frac{W_{abs}}{I} \qquad (\text{eq. 2})$$

$$\sigma_{sc} = \frac{W_{sc}}{I} \qquad (\text{eq. 3})$$

Where $W_{abs}$ and $W_{sc}$ are energy absorbed and scattered per unit time by the nanoparticle respectively, and $I$ is the intensity of incident light.

$$I = \tfrac{1}{2} c \varepsilon \ |E_{inc}|^2 \ \hat{k} \qquad (eq.\ 4)$$

Where $\hat{k}$ denotes the propagation direction of the incident light wave. The power absorbed and scattered in eq. 2 is obtained from the numerical solution as follows:

$$W_{abs} = \frac{1}{2} \iiint_V \mathrm{Re}[(\sigma E + j\omega D).E^* + j\omega B.H^*] dV \qquad (eq.\ 5)$$

$$W_{sc} = \frac{1}{2} \iint_S \mathrm{Re}[E_{sc} \times H_{sc}^*].n dS \qquad (eq.\ 6)$$

Where, $D$ and the superscript * denote displacement currents and complex conjugate respectively. In eq. 5, the integration is throughout the volume of the nanoparticle and in eq. 6 the surface integral is over the surface of the nanoparticle. These computations were carried out using built-in functions of COMSOL Multiphysics.

The heat transfer analysis was carried out by numerically solving the steady state form of heat equation:

$$U.\nabla T = \frac{k}{\rho C_p} \nabla^2 T + q \qquad (eq.\ 7)$$

Here, $T$ and $U$ are temperature and velocity vector respectively. While, $k$, $\rho$, and $C_p$ are the local properties namely, thermal conductivity, density and heat capacity. The local heat generated per unit time or the power per unit volume as a function of position is represented by $q$ which is directly obtained from $W_{abs}$. Due to extremely small length scale of the nanoparticle clusters, also implying an extremely small Grashof number, the convective contribution to heat transfer can be neglected and the left-hand side term in eq. 7 can be removed. Thus, the problem conveniently simplifies to solving the conduction equation in a suitably large domain that imitates an infinite medium. The thermophysical properties of Au and water are taken from Chen *et al.*[10] Considering the strong variation of the thermal conductivity of Au as a function of the thickness of the film used for measurement, the temperatures for $k = 110$ W/m.K and $k = 317$ W/m.K were compared to show that the variability of the thermal conductivity has little effect on the steady state temperature profile (Table S1 and Figure S2). Since the thermal conductivity of water is significantly low as compared to Au, this variability does not impact the conductive heat transfer characteristics. From dimensional considerations, for both the values of $k$, the value of Biot number is very low implying that lump capacitance approximation is valid for both the cases.[59]

## 3. Results and discussion

The reliability of the present computational method is established upon previous literature that has shown good agreement between COMSOL Multiphysics numerical solutions, analytical solutions, as well as with experimental results.[60-61] Even for more complex nanostructures, good agreement between the computational method and experimental data has been obtained. [40,62,8] For the present work, Figure S1 and S2 illustrates the validation of the numerical models used in this study by comparing them to existing numerical as well as experimental results, section 3.1.[62,58] Since the photothermal effect is direct consequence of the optical characteristics of the nanostructures considered, the optical behavior is discussed in details in section 3.2, followed by the discussion on the thermal effects of the important cases in section 3.3.

### 3.1 Model validation

Reproduction of the results by Fan *et al*.[62] and Barrow *et al*.[58] are useful as their experimental results for Au nano-shell heptamers and nanoparticle trimers, respectively, are in excellent agreement with numerical results (Figure S1). Fan *et al*. also showed experimentally that the effect of a silica substrate on the nano-shell heptamer is weak. Despite using the same numerical framework, the minor deviation of present results from the numerical results of Barrow *et al*. can be attributed to the mesh quality, PML or domain size. Nonetheless, the spectral features are in complete agreement. Importantly, the agreement between experimental and theoretical results from classical electromagnetic theory when the inter-particle gap is as small as 0.5 to 1 nm validates the exclusion of tunneling effects. Additionally, the results reported by Baffou *et al.* for an array of 15 nanospheres were also compared with present results to compare BEM and FEM based computations (Figure S2). At this point, it is concluded that the model built in COMSOL Multiphysics is successful in reproducing various sets of experimental and numerical data from literature, and will from hereon be used to study clustering effects in more detail.

As even small clusters with 3 equilaterally placed nanoparticles exhibit polarization independence for in-plane polarized incident light, it is naturally expected that bigger clusters will also exhibit similar planar isotropy and the results in Figure 2(a) illustrates this for a 37 nanoparticle cluster. To understand this polarization independence *i.e.* axial symmetry, it is more convenient to work with a simple system like an equilateral trimer for which, using plasmon hybridization theory, nine hybrid plasmon modes (or eigen modes) can be derived.[22,63] The excitable planer modes for an equilateral trimer are symmetric linear combination of individual plasmon modes meaning, plasmons mix into each other with the same symmetries that they exhibit at infinite separation.[64] Thus, the symmetry in a larger cluster can be explained on the same principle. However, such an analysis for a 37 nanoparticle cluster would be much more complicated. Figure 2(b) shows for the same cluster that the system also has 2-fold symmetry in terms of individual optical spectra about the dotted axes shown in the diagram in the inset, due to which the individual spectra of corresponding nanoparticles from each quarter coincide completely. Thus, the cluster can be divided into four equal sections equivalent to one another. Figure S4 presents the absorption spectra of individual nanoparticles in a 61 nanoparticle cluster. It is clear from both Figure S4 and Figure 2 that the absorption spectra for individual nanoparticles vary strongly with the position in the cluster implying that similar position dependence of the photothermal effects can also be expected. This aspect is discussed further in the later sections addressing the near-field and thermal effects.

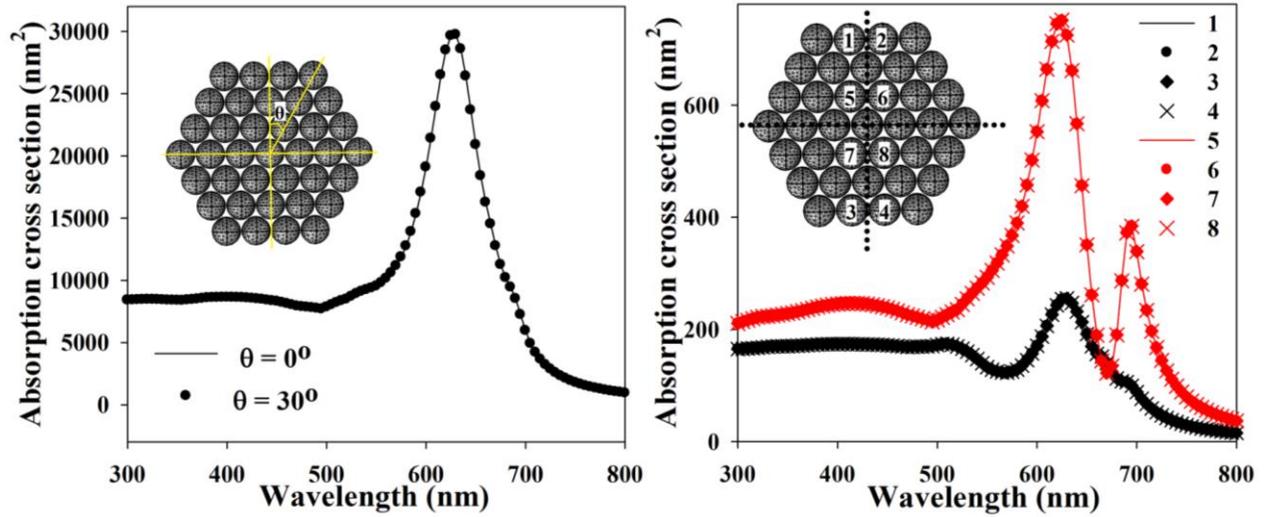

Figure 2. (a) Effect of in-plane polarization angle of incident light on the total absorption cross section (b) Comparison of absorption cross section of individual nanoparticles to show the 2-fold symmetry for polarization angle 0º.

## 3.1 Effect of cluster size for strongly interacting nanoparticles

In order to delineate the effect of cluster size, spectra for closed packed nanoparticle clusters were compared by varying the number of nanospheres from 1 to 61. Since the individual radiative damping is very small for 20 nm nanoparticles, the inter-particle distance considered was 1 nm so that there is strong interaction among the near-fields of individual nanoparticle. As apparent from Figure 3, the LSPR is red shifting for both absorption and scattering spectra with increasing number of nanoparticles. Obviously, increasing the number of particles will increase the total optical intensity. However, from efficiency perspective, it is more important to quantify the intensity as the average intensity per particle. Thus, the comparison in Figure 3 is shown in terms of average absorption and scattering cross section per nanoparticle. It is rather intuitive to expect that in the limit of an infinite number of nanoparticles, the average optical cross section per nanoparticle will remain unchanged upon further expansion of the cluster. This is quite clear for the average absorption cross section while varying the cluster size from 1 to 61 nanoparticles in Figure 3. It is worth noting that in the infinity limit, it is the average optical intensity, absorption or scattering, that remains constant upon addition of nanoparticles as the total absorption and scattering will obviously increase with more nanoparticles. While, although the average scattering cross section continually increases till 61 nanoparticles, the rate of increase becomes smaller implying that at a certain number it will reach a plateau or a maximum. Thus, the optimal cluster size is not the same for absorption and scattering. Interestingly, 20 nm nanospheres have a significantly large average scattering cross section per particle in the cluster in contrast to their diminishingly small scattering cross section when isolated (Figure 3b). Since an isolated 20 nm nanoparticle is essentially "non-scattering", this enhancement in scattering can be attributed to near-field coupling of the nanoparticles.[15,65] If the collective optical response of the nanoparticles in a cluster is due to the interaction of near-fields, the cluster can behave like a single entity bridged together by inter-connected near-fields (see further, Figure 7). Since the radiative modes of individual particle become strong in cluster, the coupling of these strengthened radiative modes is also possible. Thus, the collective response is a correlative effect of near-field and far-field interactions. However, the relative strength of these two interaction modes is an important aspect discussed later in this work, Figure 5, 6 and 7. Another interesting way to view this collective behavior is a hybridized plasmon mode of constituent individual plasmon modes analogous to a molecular orbital from the hybridization of individual atomic orbitals in molecular orbital theory.[29,66,67] Interestingly, akin to the red-shift of optical cross sections with increasing size for an isolated nanostructure, the collective optical spectra for clusters also red-shift with increasing cluster size. Also, the scattering is greatly enhanced for bigger clusters. An important feature of a nanoparticle cluster is that as the cluster size increases, even to the macro-scale, the localized plasmons remain intact. In contrast, a single nanoparticle is known to lose its optical properties derived from localized surface plasmon resonance if the size increases beyond a certain point, as the localized plasmons give way to bulk plasmons. The LSPR wavelength vs. number of nanoparticles in the cluster is plotted in Figure 4 and clearly evidences that the red shift tends to become weaker and weaker with increasing cluster size. Thus, after a certain limit, there will not be any red-shift upon addition of more nanoparticles implying that the infinite limit is reached. While extending the present numerical model to find this infinite limit is a coveted aspect, the increase in the required computational power with increasing cluster size unfortunately limits this endeavor. Thus, the use of theoretical formalisms of Modinos *et al*. or KKR method will be more practical in this regard.[33-37]

The coupled dipole method is an alternative approach to tackle this problem, but it is based upon assumptions that significantly simplify the model.[38,68] Use of a theoretical formalism for infinite arrays of nanoparticle remains as an important objective of our future research.

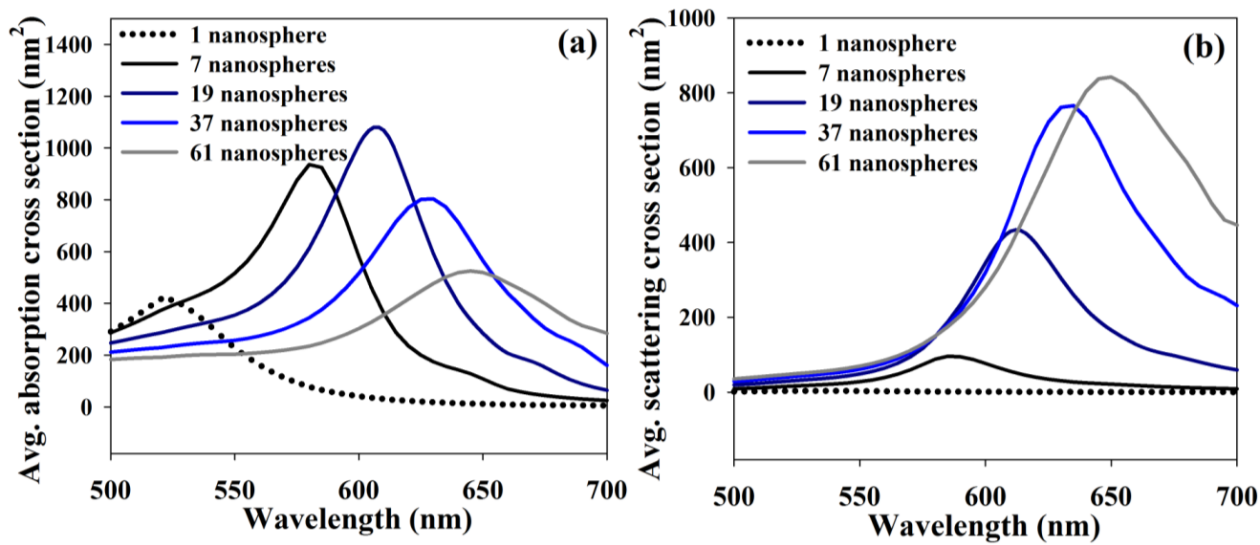

Figure 3. (a) Average absorption cross section per nanoparticle (b) average scattering cross section per nanoparticle for varying cluster size with inter-particle gap of 1 nm.

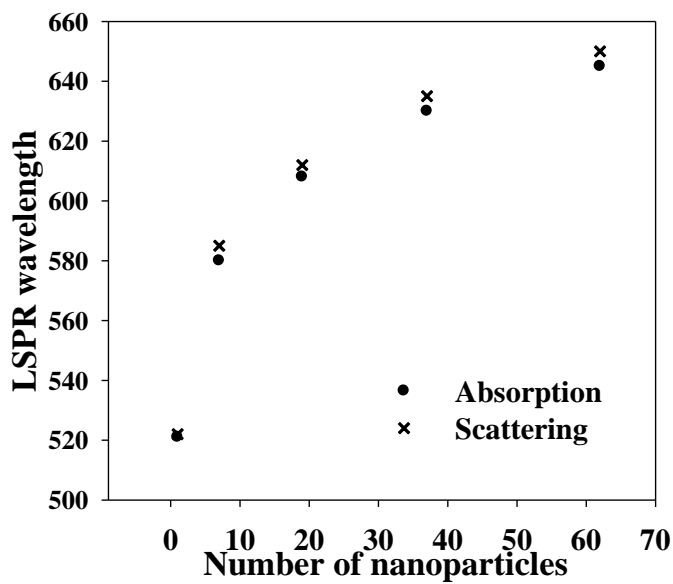

Figure 4. Absorption and scattering spectral peak wavelength variation with number of nanoparticles in the cluster.

**3.2 Effect of inter-particle gap and particle size**

As discussed above, Figure 3 (a) and (b) show that the absorption per particle increases (then decreases) with cluster size and otherwise non-radiative 20 nm nanoparticles have significant 'per particle' average scattering cross sections when they are clustered together. To further understand the nature of this interaction, it is instructive to study the effect of inter-particle distance in a cluster with particles with different individual scattering power.

To start with, the effect of inter-particle distance was studied for a 19 nanoparticle cluster, as it has been shown for closely packed ($d$ = 1 nm) 20 nm nanoparticles that the average absorption cross section per nanoparticle is the highest in that case. Now in Figure 5 (a) and (b), both absorption and scattering spectra red-shift with increasing intensity as the inter-particle distance decreases. These trends are consistent with the hypothesis that the inter-particle interaction happens primarily through overlapping near-fields as the spectral features are intact. Thus, with decreasing gap, the cluster tends to show a collective optical response like an integrated large nanostructure. The inter-particle interaction in clusters is understood further by analyzing the effect of the inter-particle gap in a cluster of 40 nm nanoparticles for which the individual scattering power is higher (Figure 6). The interaction among these bigger nanoparticles is enabled through both overlapping of the near-fields *i.e.*, near-field coupling and the interference of the radiative modes *i.e.*, far-field coupling. Importantly, when 40 nm nanoparticles are clustered, the characteristic resonance peak in the absorption spectra loses its sharp feature and intensity, while the scattering intensity increases. The evidence of the strong interference of the radiative modes is the fano-like characteristic resonance in the scattering spectra resulting from superradiant and subradiant modes for an inter-particle gap of 2 and 4 nm.[69] It is interesting to note here that the fano-dip is also dependent on the inter-particle gap. This intensification of the radiative damping through the interference of scattered radiation may be connected to the loss of spectral characteristics in the absorption spectra. Evidently, the larger nanoparticles require a larger inter-particle distance to be free from inter-particle interaction. Figure 7 shows the difference in the strength of inter-particle interaction for a 1 nm and a 16 nm inter-particle gap by comparing the near-fields. As expected for the 1 nm gap, the strength of the electric field between the particles is significantly higher than that for a 16 nm gap. Also, for 16 nm gap the dipolar plasmon modes of individual nanoparticles are dominant with weak enhancement of the near-field in the inter-particle gap. In contrast, for the 1 nm gap individual dipolar modes are lost as they couple together to form hybridized modes. This hybridized modes results in strong near-field enhancement in specific locations of the cluster.

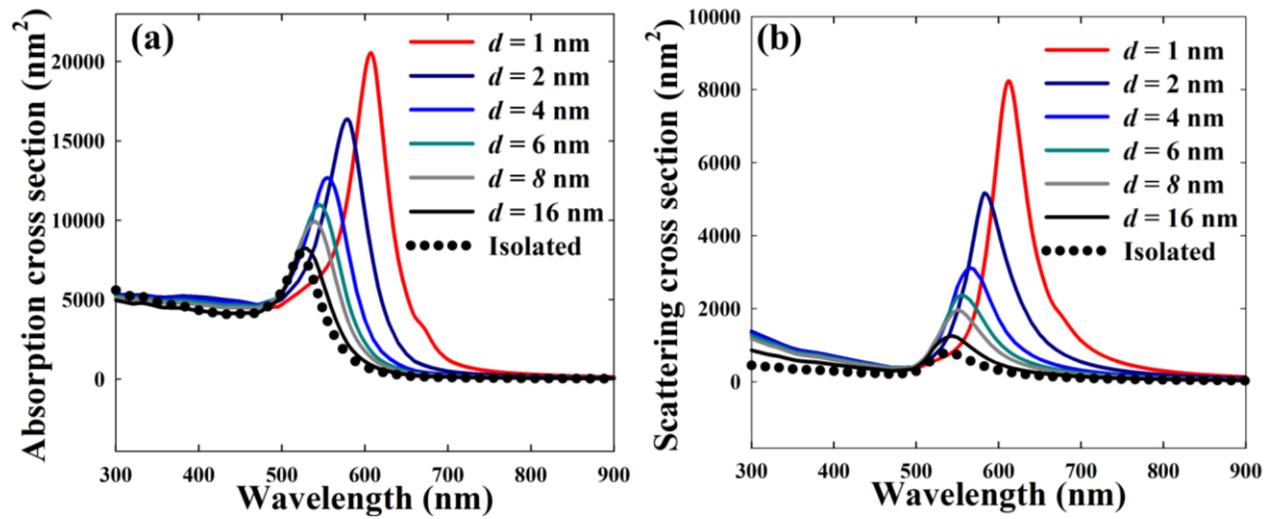

Figure 5. Variation of (a) absorption and (b) scattering cross section of a 19 nanoparticle cluster with interparticle distance for nanoparticle size 20 nm.

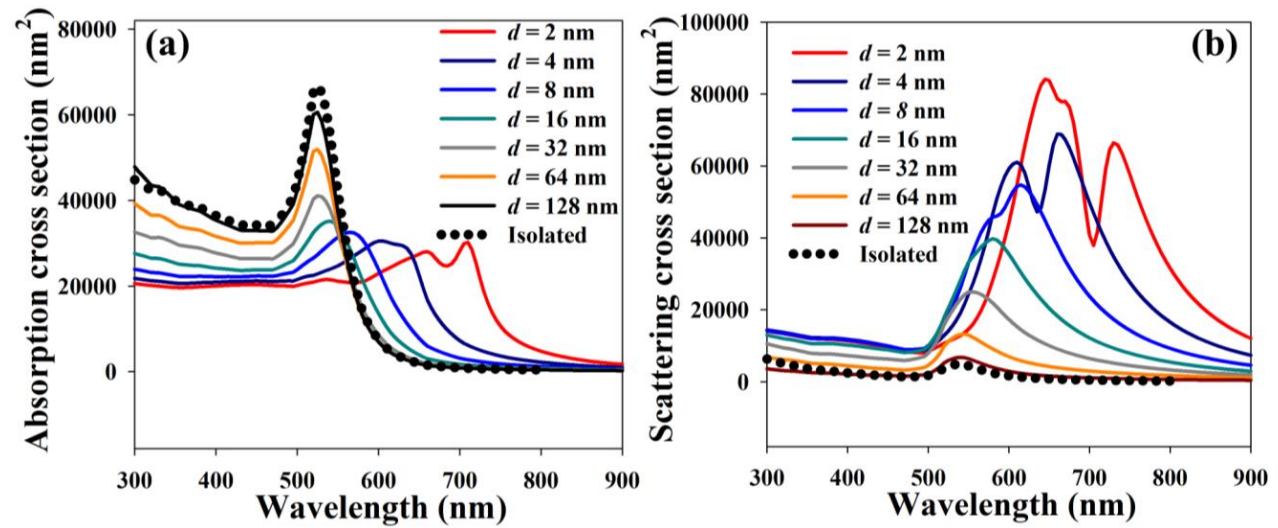

Figure 6. Variation of (a) absorption and (b) scattering cross section of a 19 nanoparticle cluster with inter-particle distance for nanoparticle size 40 nm.

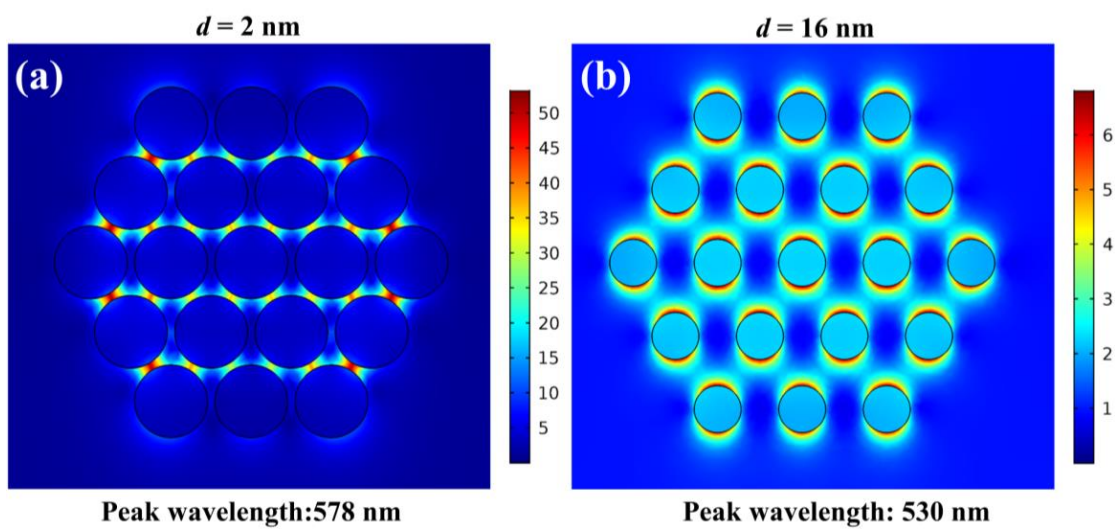

Figure 7. Near-field enhancement in 19 nanoparticle clusters with inter-particle distance (a) 2 nm and (b) 16 nm for a nanoparticle size of 20 nm.

The near field enhancement at specific locations in a cluster for a certain incident wavelength has direct correspondence with the fact that individual nanoparticles in a cluster have significantly different absorption characteristics. Apparently in Figure S4, the absorption spectra of individual nanoparticles in a 61 nanoparticle cluster differ one to another greatly with varying peak positions in the spectrum. Thus, the near-field enhancement corresponding to these individual absorption peaks also varies with particle location. Figure S5 shows the near-field enhancement in a 61 nanoparticle cluster for different incident wavelengths. For 645 nm incident wavelength, the near-field enhancement is along the nanoparticles at the left and right boundaries of the cluster. But at 720 nm wavelength, the near-field enhancement is now in the core of the cluster. This spatial variability of near-field enhancement is an interesting observation with regard to many applications of plasmonics. For instance, it has been found that strong near-field enhancement and hot-electron injection catalyzes certain reactions, photocatalytic processes, etc.[2,4,5] Local control of these processes is possible due to this spatial variability of near-field enhancement.

## 3.3 Photothermal effect of clusters

The inevitable consequence of energy absorption is thermal dissipation and thus, temperature rise of the nanoparticle cluster and its surrounding, while the scattered energy is simply radiated away. Thus, the absorption spectra of the nanoparticles and their assemblies have direct correspondence with the resulting temperature fields. Since the absorption characteristics are dependent on the nanoparticle position in a cluster, individual nanoparticles will exhibit different steady state temperatures according to their relative position and polarization angle, Figure 8 (a) to (d). Interestingly, the number of nanoparticles strongly influences the steady state temperature of the cluster due to enhanced absorption in hybridized plasmon modes in a cluster and containment of heat by the outer spheres of the cluster. Thus, absorption and heat containment simultaneously influences the steady state temperatures of the particles. In the water medium at 25 °C, a single nanosphere is heated up to only ~30 °C (Figure S6), while the 7 nanoparticle cluster attains a temperature over 62 °C when irradiated at 1 mW/$\mu m^2$. Similarly, for 19 and 37 nanoparticle clusters, the temperature of the "hottest" nanoparticle in the cluster will exceed 100 °C. Interestingly, the average per particle absorption is higher for 19 nanoparticle cluster as compared to the 37 nanoparticle cluster, but the 37 nanoparticle cluster has a higher steady state temperature for the central nanoparticles than the 19 nanoparticle cluster. Clearly, although the absorption is less, the central nanoparticles in the 37 nanoparticle cluster are surrounded by more nanoparticles than a 19 nanoparticle cluster implying greater heat retention and steady state temperature in the core. However, for the 61 nanoparticle cluster, the steady state temperature of the "hottest" particles again drops to ~95 °C because the decrease in the average absorption per particle is the dominating factor. Also for the 61 nanoparticle cluster, nanoparticles at the boundaries have high temperature in spite of lower heat retention as the high absorption compensates for the higher heat loss rate to the surrounding. The temperature variation with the position of the particles is an important aspect demonstrated here. In Figure 8(e), the change of the temperature profile in the 37 nanoparticle cluster is shown when the incident radiation polarization is horizontal is shown in contrast to vertical polarization in Figure 8(c). This indicates the possibility of controlling the temperature profile in a cluster. The temperature field of the 61 nanoparticle cluster with the surrounding is shown in Figure 8(f).

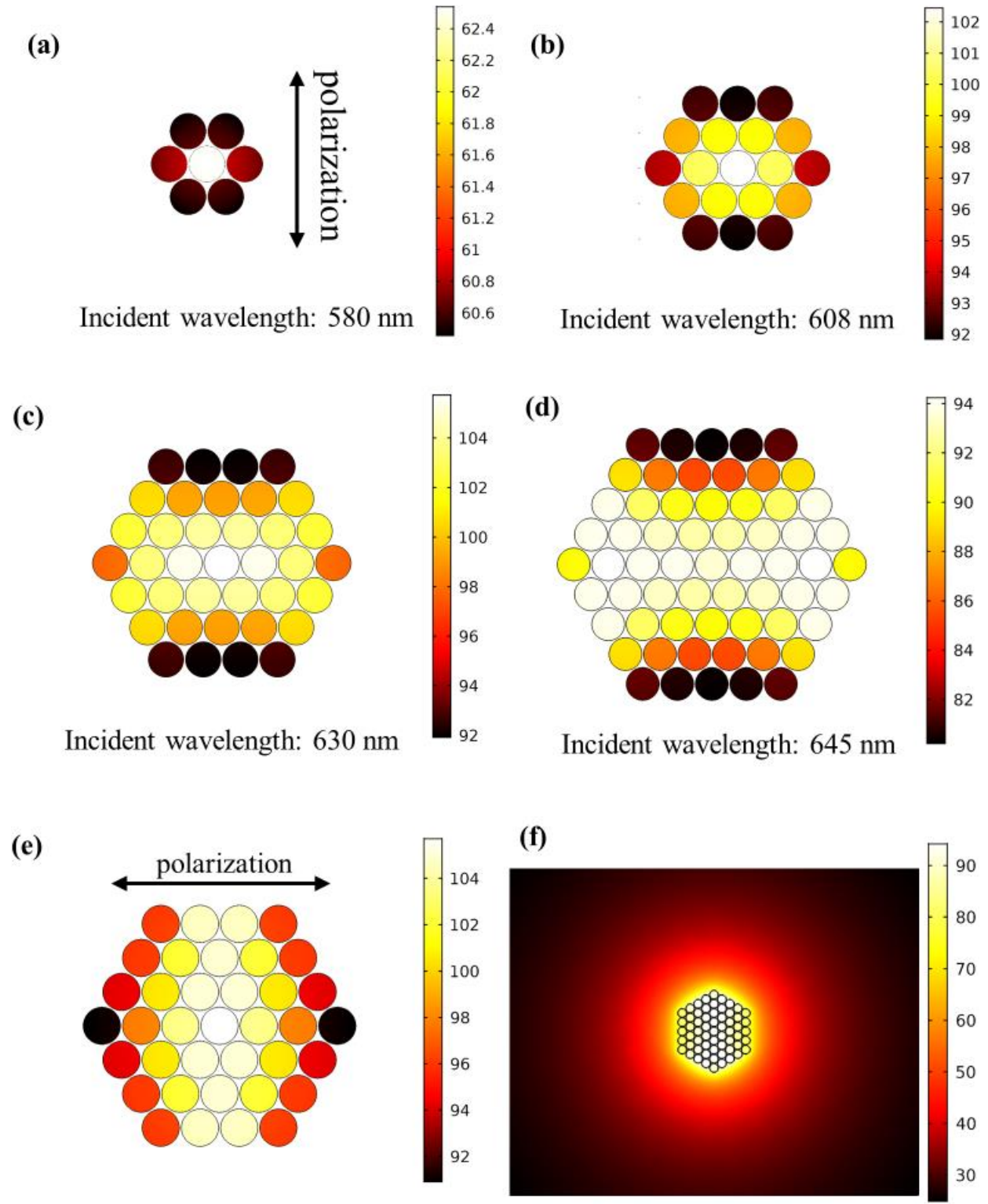

Figure 8. Steady state temperatures (ºC) of nanoparticles in a cluster: (a) 7 nanoparticles (b) 19 nanoparticles (c) 37 nanoparticles (d) 61 nanoparticles upon EM irradiation of respective peak wavelength in infinite water medium. (e) Temperature profile in a 37 nanoparticle cluster when the incident radiation polarization is horizontal. (f) Temperature field around a 61 nanoparticle cluster. The intensity of incident radiation is 1 mW/μm$^2$ for all the cases.

It is worth mentioning that although the steady state temperature is significantly enhanced by clustering nanoparticles as compared to isolated nanoparticles, the light intensity required to achieve this heating effect is considerably high. In Figure S7, it is clear that for a drop of one order of magnitude in the intensity results in a significantly lower steady state temperature. When the intensity is two orders of magnitude lower, there is almost no apparent heating. However, it is important to note that the present model assumes an infinite water medium which is not an insulating material. To check the extreme limit, heat transfer analysis was also done for a 37 nanoparticle cluster in an air medium, a good insulating medium, while keeping the absorption values the same as those in water, Figure S8. For the intensity corresponding to Figure 9, the overall cluster temperature was estimated to be as high as around 900 ºC, attained solely by altering the thermal properties of the medium. It is to be noted that the convective effects are excluded in this calculation. With this fundamental insight, it is clear that raising the steady state temperature is a matter of improving the energy input rate and heat retention by better optical and thermal design of the system. For instance, attaining a high temperature under sunlight would require improved heat trapping possibly by clustering in 3D and embedding in an insulating matrix as shown by Dhiman *et al*. for dendritic plasmonic colloidosomes.[70] In contrast to 20 nm nanoparticles considered in this work for which the heating is rather weak, an isolated large nanoparticle of 100 nm diameter can induce a temperature rise >50 ºC for the intensity considered in this study.[71] This promises further development in the cluster heating with larger nanoparticles. The present results demonstrate the possibilities to improve and control the thermal effects of nanoparticles by clustering in suitable configurations.

The effect of inter-particle distance on the absorption spectra of clusters is another aspect that leads to interesting thermal effects. In Figure 8(b), it has been shown that a 19 nanoparticle cluster can attain significantly high temperature as compared to an isolated nanoparticle when irradiated with electromagnetic wave of their peak wavelengths. However, this temperature is for the smallest inter-particle gap of 1 nm. As the inter-particle gap increases, the absorption intensity as well as the spectral peak changes. The consequence of this on the thermal effect is obviously direct. In Figure 9, the steady state temperature of a 19 nanoparticle cluster with varying inter-particle gap is shown. For an inter-particle gap of 8 nm, the thermal effects of the nanoparticle cluster are practically gone. It is however useful to point out the difference in the representation of Figure 8 and 9. In Figure 8, the spatial variation of temperature only in the nanoparticles is shown, while in Figure 9, the cluster as well as the surrounding temperature field is shown. The results in Figure 9 also promises interesting applications of local temperature control and sensing. For instance, it has been shown how the inter-particle gap of nanoparticle assembly embedded in a substrate can be controlled by thermal treatment, chemical modification or mechanical strain of the substrate without structural distortion.[72-74] A visual demonstration of this effect can be found in the video attached as supporting document. It is, however, important to note that demonstrated photothermal effect due to inter-particle gap is only possible with 20 nm particles according to our simulation. As we have shown that the 40 nm nanoparticle cluster does not exhibit similar dependence of absorption characteristics on the inter-particle gap. Due to strong coupling of radiative modes, the absorption rather decreases with smaller inter-particle gap. Detailed experimental studies on 2D clusters of weakly radiant nanoparticles as small as 20 nm have not been carried out yet. Thus, experimental verification of these theoretical results is of significant practical importance with regard to the numerous possible applications.

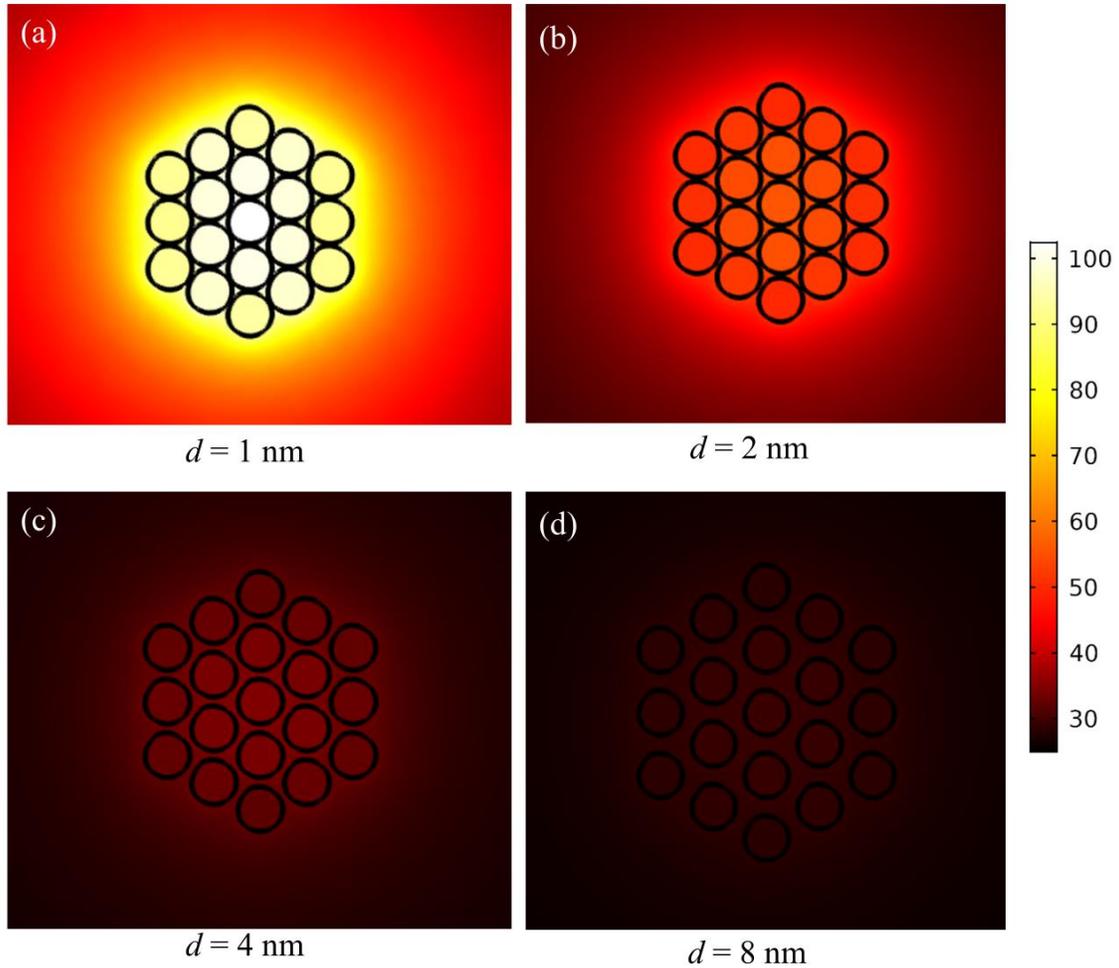

Figure 9. Steady state temperatures (ºC) of a 19 nanoparticle cluster with varying inter-particle distance, $d$. The wavelength of the incident radiation is the peak wavelength of 608 nm for $d = 1$ nm. The intensity is 1 mW/µm$^2$ for all the cases.

## 4. Conclusion

The present study investigates the collective plasmonic characteristics and resulting photothermal effects of isolated Au nanoparticle clusters of different sizes. The collective optical response is analyzed in terms of near-field and far-field coupling among constituent nanoparticles in the cluster. Otherwise weakly radiative 20 nm Au nanoparticles, when coupled through overlapping of near fields in a closed packed 2D cluster, exhibit optical response as an integrated structure resulting in strong absorption and scattering with a spectral shift. As optical cross sections obviously increase with increasing cluster size, it is more useful to determine average absorption/scattering per particle from efficiency perspective. As shown for 20 nm nanoparticle clusters with small inter-particle gap (1 nm) i.e. strong near-field coupling, the absorption intensity per particle increases with cluster size up to an optimal cluster size beyond which the average intensity decreases to reach a constant value at the infinity limit. For absorption, the optimal cluster size is 19 nanoparticles, while, for scattering the trend could not be evident in the present results that covers up to a maximum cluster size of 61 nanoparticles. The spectral shift with increasing cluster size is towards red. Since the coupling among weakly scattering 20 nm nanoparticles is predominantly through the near-fields, increasing inter-particle gap rapidly weakens the collective modes and the individual dipolar modes of the nanoparticles become strong. However, due to their stronger radiative modes, 40 nm nanoparticles in cluster are coupled through both near-field and far-field enabling far reaching inter-particle interaction. The effect of decreasing inter-particle gap i.e. stronger collective modes in a cluster is red shift of the spectra. However, the contrast between 20 nm and 40 nm nanoparticles in the context of the effect of inter-particle gap is that the absorption intensity increases for 20 nm particles while decreases for 40 nm particles with decreasing gap. Also, the absorption characteristics in a cluster vary strongly from particle to particle. Since, the steady state temperature of individual particles depends upon the individual absorption intensity and the heat retention around the particle, the steady state temperature of individual nanoparticles vary strongly in a cluster. This variation is stronger for stronger laser intensity and also dependent on the incident polarization. The effect of inter-particle distance on absorption intensity leads to interesting photothermal effects in a cluster which can be controlled by varying this distance. The present study provides new theoretical insight into plasmonic behavior of closed packed Au nanoparticle clusters and resulting photothermal phenomena. The photothermal effect in the clusters shows possibility of nanoscale temperature control. This understanding will be useful in the implementation of 2D nanoparticle clusters in relevant applications.

**Supporting information**

Model validation, sensitivity analysis, additional results on optical spectra and thermal effects.

**Acknowledgment**

The financial assistance from the University of Antwerp for this work is duly acknowledged.

**References**

(1) Gramotnev, D. K.; Bozhevolnyi, S. I. Plasmonics beyond the Diffraction Limit. *Nat. Photonics* **2010**, *4* (2), 83–91. https://doi.org/10.1038/nphoton.2009.282.

(2) Verbruggen, S. W. TiO2 Photocatalysis for the Degradation of Pollutants in Gas Phase: From Morphological Design to Plasmonic Enhancement. *J. Photochem. Photobiol. C Photochem. Rev.* **2015**, *24*, 64–82. https://doi.org/10.1016/j.jphotochemrev.2015.07.001.

(3) Verbruggen, S. W.; Keulemans, M.; Goris, B.; Blommaerts, N.; Bals, S.; Martens, J. A.; Lenaerts, S. Plasmonic 'Rainbow' Photocatalyst with Broadband Solar Light Response for Environmental Applications. *Appl. Catal. B Environ.* **2016**, *188*, 147–153. https://doi.org/10.1016/j.apcatb.2016.02.002.

(4) Asapu, R.; Claes, N.; Ciocarlan, R.-G.; Minjauw, M.; Detavernier, C.; Cool, P.; Bals, S.; Verbruggen, S. W. Electron Transfer and Near-Field Mechanisms in Plasmonic Gold-Nanoparticle-Modified TiO2 Photocatalytic Systems. *ACS Appl. Nano Mater.* **2019**, *2* (7), 4067–4074. https://doi.org/10.1021/acsanm.9b00485.

(5) Stewart, M. E.; Anderton, C. R.; Thompson, L. B.; Maria, J.; Gray, S. K.; Rogers, J. A.; Nuzzo, R. G. Nanostructured Plasmonic Sensors. *Chem. Rev.* **2008**, *108* (2), 494–521. https://doi.org/10.1021/cr068126n.

(6) Boriskina, S. V.; Ghasemi, H.; Chen, G. Plasmonic Materials for Energy: From Physics to Applications. *Mater. Today* **2013**, *16* (10), 375–386. https://doi.org/10.1016/j.mattod.2013.09.003.

(7) Willets, K. A.; Wilson, A. J.; Sundaresan, V.; Joshi, P. B. Super-Resolution Imaging and Plasmonics. *Chem. Rev.* **2017**, *117* (11), 7538–7582. https://doi.org/10.1021/acs.chemrev.6b00547.

(8) Jiang, N.; Zhuo, X.; Wang, J. Active Plasmonics: Principles, Structures, and Applications. *Chem. Rev.* **2018**, *118* (6), 3054–3099. https://doi.org/10.1021/acs.chemrev.7b00252.

(9) Liao, T.-W.; Verbruggen, S. W.; Claes, N.; Yadav, A.; Grandjean, D.; Bals, S.; Lievens, P. TiO2 Films Modified with Au Nanoclusters as Self-Cleaning Surfaces under Visible Light. *Nanomaterials* **2018**, *8* (1), 30. https://doi.org/10.3390/nano8010030.

(10) Chen, X.; Chen, Y.; Yan, M.; Qiu, M. Nanosecond Photothermal Effects in Plasmonic Nanostructures. *ACS Nano* **2012**, *6* (3), 2550–2557. https://doi.org/10.1021/nn2050032.

(11) Lu, X.; Rycenga, M.; Skrabalak, S. E.; Wiley, B.; Xia, Y. Chemical Synthesis of Novel Plasmonic Nanoparticles. *Annu. Rev. Phys. Chem.* **2009**, *60* (1), 167–192. https://doi.org/10.1146/annurev.physchem.040808.090434.

(12) Verbruggen, S. W.; Keulemans, M.; Martens, J. A.; Lenaerts, S. Predicting the Surface Plasmon Resonance Wavelength of Gold–Silver Alloy Nanoparticles. *J. Phys. Chem. C* **2013**, *117* (37), 19142–19145. https://doi.org/10.1021/jp4070856.

(13) Chatterjee, H.; Rahman, D. S.; Sengupta, M.; Ghosh, S. K. Gold Nanostars in Plasmonic Photothermal Therapy: The Role of Tip Heads in the Thermoplasmonic Landscape. *J. Phys. Chem. C* **2018**, *122* (24), 13082–13094. https://doi.org/10.1021/acs.jpcc.8b00388.

(14) Wang, W.; Ramezani, M.; Väkeväinen, A. I.; Törmä, P.; Rivas, J. G.; Odom, T. W. The Rich Photonic World of Plasmonic Nanoparticle Arrays. *Mater. Today* **2018**, *21* (3), 303–314. https://doi.org/10.1016/j.mattod.2017.09.002.

(15) Kravets, V. G.; Kabashin, A. V.; Barnes, W. L.; Grigorenko, A. N. Plasmonic Surface Lattice Resonances: A Review of Properties and Applications. *Chem. Rev.* **2018**, *118* (12), 5912–5951. https://doi.org/10.1021/acs.chemrev.8b00243.

(16) Alù, A.; Engheta, N. The Quest for Magnetic Plasmons at Optical Frequencies. *Opt. Express* **2009**, *17* (7), 5723–5730. https://doi.org/10.1364/OE.17.005723.

(17) Mirin, N. A.; Bao, K.; Nordlander, P. Fano Resonances in Plasmonic Nanoparticle Aggregates. *J. Phys. Chem. A* **2009**, *113* (16), 4028–4034. https://doi.org/10.1021/jp810411q.

(18) Punj, D.; Regmi, R.; Devilez, A.; Plauchu, R.; Moparthi, S. B.; Stout, B.; Bonod, N.; Rigneault, H.; Wenger, J. Self-Assembled Nanoparticle Dimer Antennas for Plasmonic-Enhanced Single-Molecule Fluorescence Detection at Micromolar Concentrations. *ACS Photonics* **2015**, *2* (8), 1099–1107. https://doi.org/10.1021/acsphotonics.5b00152.

(19) Kuttner, C. Plasmonics in Sensing: From Colorimetry to SERS Analytics. *Plasmonics* **2018**. https://doi.org/10.5772/intechopen.79055.

(20) Luk'yanchuk, B.; Zheludev, N. I.; Maier, S. A.; Halas, N. J.; Nordlander, P.; Giessen, H.; Chong, C. T. The Fano Resonance in Plasmonic Nanostructures and Metamaterials. *Nat. Mater.* **2010**, *9* (9), 707–715. https://doi.org/10.1038/nmat2810.

(21) Hopkins, B.; Poddubny, A. N.; Miroshnichenko, A. E.; Kivshar, Y. S. Revisiting the Physics of Fano Resonances for Nanoparticle Oligomers. *Phys. Rev. A* **2013**, *88* (5), 053819. https://doi.org/10.1103/PhysRevA.88.053819.

(22) Zohar, N.; Chuntonov, L.; Haran, G. The Simplest Plasmonic Molecules: Metal Nanoparticle Dimers and Trimers. *J. Photochem. Photobiol. C Photochem. Rev.* **2014**, *21*, 26–39. https://doi.org/10.1016/j.jphotochemrev.2014.10.002.

(23) Klinkova, A.; Choueiri, R. M.; Kumacheva, E. Self-Assembled Plasmonic Nanostructures. *Chem. Soc. Rev.* **2014**, *43* (11), 3976–3991. https://doi.org/10.1039/C3CS60341E.

(24) Zhang, Y.; Wen, F.; Zhen, Y.-R.; Nordlander, P.; Halas, N. J. Coherent Fano Resonances in a Plasmonic Nanocluster Enhance Optical Four-Wave Mixing. *Proc. Natl. Acad. Sci.* **2013**, *110* (23), 9215–9219. https://doi.org/10.1073/pnas.1220304110.

(25) Dahmen, C.; Schmidt, B.; von Plessen, G. Radiation Damping in Metal Nanoparticle Pairs. *Nano Lett.* **2007**, *7* (2), 318–322. https://doi.org/10.1021/nl062377u.

(26) Hao, E.; Schatz, G. C. Electromagnetic Fields around Silver Nanoparticles and Dimers. *J. Chem. Phys.* **2003**, *120* (1), 357–366. https://doi.org/10.1063/1.1629280.

(27) Zhu, W.; Esteban, R.; Borisov, A. G.; Baumberg, J. J.; Nordlander, P.; Lezec, H. J.; Aizpurua, J.; Crozier, K. B. Quantum Mechanical Effects in Plasmonic Structures with Subnanometre Gaps. *Nat. Commun.* **2016**, *7*, 11495. https://doi.org/10.1038/ncomms11495.

(28) Asapu, R.; Ciocarlan, R.-G.; Claes, N.; Blommaerts, N.; Minjauw, M.; Ahmad, T.; Dendooven, J.; Cool, P.; Bals, S.; Denys, S.; et al. Plasmonic Near-Field Localization of Silver Core–Shell Nanoparticle Assemblies via Wet Chemistry Nanogap Engineering. *ACS Appl. Mater. Interfaces* **2017**, *9* (47), 41577–41585. https://doi.org/10.1021/acsami.7b13965.

(29) Nordlander, P.; Oubre, C.; Prodan, E.; Li, K.; Stockman, M. I. Plasmon Hybridization in Nanoparticle Dimers. *Nano Lett.* **2004**, *4* (5), 899–903. https://doi.org/10.1021/nl049681c.

(30) Baur, S.; Sanders, S.; Manjavacas, A. Hybridization of Lattice Resonances. *ACS Nano* **2018**, *12* (2), 1618–1629. https://doi.org/10.1021/acsnano.7b08206.

(31) Tira, C.; Tira, D.; Simon, T.; Astilean, S. Finite-Difference Time-Domain (FDTD) Design of Gold Nanoparticle Chains with Specific Surface Plasmon Resonance. *J. Mol. Struct.* **2014**, *1072*, 137–143. https://doi.org/10.1016/j.molstruc.2014.04.086.

(32) Zhong, Z.; Patskovskyy, S.; Bouvrette, P.; Luong, J. H. T.; Gedanken, A. The Surface Chemistry of Au Colloids and Their Interactions with Functional Amino Acids. *J. Phys. Chem. B* **2004**, *108* (13), 4046–4052. https://doi.org/10.1021/jp037056a.

(33) Modinos, A. Scattering of Electromagnetic Waves by a Plane of Spheres-Formalism. *Phys. Stat. Mech. Its Appl.* **1987**, *141* (2), 575–588. https://doi.org/10.1016/0378-4371(87)90184-1.
(34) Stafanou, N.; Modinos, A. Scattering of Light from a Two-Dimensional Array of Spherical Particles on a Substrate. *J. Phys. Condens. Matter* **1991**, *3* (41), 8135–8148. https://doi.org/10.1088/0953-8984/3/41/012.
(35) Stefanou, N.; Yannopapas, V.; Modinos, A. MULTEM 2: A New Version of the Program for Transmission and Band-Structure Calculations of Photonic Crystals. *Comput. Phys. Commun.* **2000**, *132* (1), 189–196. https://doi.org/10.1016/S0010-4655(00)00131-4.
(36) Moroz, A.; Sommers, C. Photonic Band Gaps of Three-Dimensional Face-Centred Cubic Lattices. *J. Phys. Condens. Matter* **1999**, *11* (4), 997–1008. https://doi.org/10.1088/0953-8984/11/4/007.
(37) Moroz, A. Metallo-Dielectric Diamond and Zinc-Blende Photonic Crystals. *Phys. Rev. B* **2002**, *66* (11), 115109. https://doi.org/10.1103/PhysRevB.66.115109.
(38) Zundel, L.; Manjavacas, A. Finite-Size Effects on Periodic Arrays of Nanostructures. *J. Phys. Photonics* **2018**, *1* (1), 015004. https://doi.org/10.1088/2515-7647/aae8a2.
(39) Liu, N.; Mukherjee, S.; Bao, K.; Li, Y.; Brown, L. V.; Nordlander, P.; Halas, N. J. Manipulating Magnetic Plasmon Propagation in Metallic Nanocluster Networks. *ACS Nano* **2012**, *6* (6), 5482–5488. https://doi.org/10.1021/nn301393x.
(40) Cui, Y.; Zhou, J.; Tamma, V. A.; Park, W. Dynamic Tuning and Symmetry Lowering of Fano Resonance in Plasmonic Nanostructure. *ACS Nano* **2012**, *6* (3), 2385–2393. https://doi.org/10.1021/nn204647b.
(41) Pazos-Perez, N.; Wagner, C. S.; Romo-Herrera, J. M.; Liz-Marzán, L. M.; García de Abajo, F. J.; Wittemann, A.; Fery, A.; Alvarez-Puebla, R. A. Organized Plasmonic Clusters with High Coordination Number and Extraordinary Enhancement in Surface-Enhanced Raman Scattering (SERS). *Angew. Chem. Int. Ed.* **2012**, *51* (51), 12688–12693. https://doi.org/10.1002/anie.201207019.
(42) Matricardi, C.; Hanske, C.; Garcia-Pomar, J. L.; Langer, J.; Mihi, A.; Liz-Marzán, L. M. Gold Nanoparticle Plasmonic Superlattices as Surface-Enhanced Raman Spectroscopy Substrates. *ACS Nano* **2018**, *12* (8), 8531–8539. https://doi.org/10.1021/acsnano.8b04073.
(43) Greybush, N. J.; Liberal, I.; Malassis, L.; Kikkawa, J. M.; Engheta, N.; Murray, C. B.; Kagan, C. R. Plasmon Resonances in Self-Assembled Two-Dimensional Au Nanocrystal Metamolecules. *ACS Nano* **2017**, *11* (3), 2917–2927. https://doi.org/10.1021/acsnano.6b08189.
(44) Le, F.; Brandl, D. W.; Urzhumov, Y. A.; Wang, H.; Kundu, J.; Halas, N. J.; Aizpurua, J.; Nordlander, P. Metallic Nanoparticle Arrays: A Common Substrate for Both Surface-Enhanced Raman Scattering and Surface-Enhanced Infrared Absorption. *ACS Nano* **2008**, *2* (4), 707–718. https://doi.org/10.1021/nn800047e.
(45) Ali, M. R. K.; Wu, Y.; El-Sayed, M. A. Gold-Nanoparticle-Assisted Plasmonic Photothermal Therapy Advances Toward Clinical Application. *J. Phys. Chem. C* **2019**, *123* (25), 15375–15393. https://doi.org/10.1021/acs.jpcc.9b01961.
(46) Baffou, G.; Ureña, E. B.; Berto, P.; Monneret, S.; Quidant, R.; Rigneault, H. Deterministic Temperature Shaping Using Plasmonic Nanoparticle Assemblies. *Nanoscale* **2014**, *6* (15), 8984–8989. https://doi.org/10.1039/C4NR01644K.
(47) Hatef, A.; Fortin-Deschênes, S.; Boulais, E.; Lesage, F.; Meunier, M. Photothermal Response of Hollow Gold Nanoshell to Laser Irradiation: Continuous Wave, Short and Ultrashort Pulse. *Int. J. Heat Mass Transf.* **2015**, *89*, 866–871. https://doi.org/10.1016/j.ijheatmasstransfer.2015.05.071.
(48) Alali, F.; Karampelas, I. H.; Kim, Y. H.; Furlani, E. P. Photonic and Thermofluidic Analysis of Colloidal Plasmonic Nanorings and Nanotori for Pulsed-Laser Photothermal Applications. *J. Phys. Chem. C* **2013**, *117* (39), 20178–20185. https://doi.org/10.1021/jp406986y.

(49) Baffou, G.; Berto, P.; Bermúdez Ureña, E.; Quidant, R.; Monneret, S.; Polleux, J.; Rigneault, H. Photoinduced Heating of Nanoparticle Arrays. *ACS Nano* **2013**, *7* (8), 6478–6488. https://doi.org/10.1021/nn401924n.

(50) Wang, S.; Fu, L.; Zhang, Y.; Wang, J.; Zhang, Z. Quantitative Evaluation and Optimization of Photothermal Bubble Generation around Overheated Nanoparticles Excited by Pulsed Lasers. *J. Phys. Chem. C* **2018**, *122* (42), 24421–24435. https://doi.org/10.1021/acs.jpcc.8b07672.

(51) Ren, Y.; Chen, Q.; Qi, H.; Ruan, L.; Dai, J. Phase Transition Induced by Localized Surface Plasmon Resonance of Nanoparticle Assemblies. *Int. J. Heat Mass Transf.* **2018**, *127*, 244–252. https://doi.org/10.1016/j.ijheatmasstransfer.2018.07.049.

(52) Baffou, G.; Quidant, R.; Girard, C. Thermoplasmonics Modeling: A Green's Function Approach. *Phys. Rev. B* **2010**, *82* (16), 165424. https://doi.org/10.1103/PhysRevB.82.165424.

(53) Baffou, G.; Polleux, J.; Rigneault, H.; Monneret, S. Super-Heating and Micro-Bubble Generation around Plasmonic Nanoparticles under Cw Illumination. *J. Phys. Chem. C* **2014**, *118* (9), 4890–4898. https://doi.org/10.1021/jp411519k.

(54) Johnson, P. B.; Christy, R. W. Optical Constants of the Noble Metals. *Phys. Rev. B* **1972**, *6* (12), 4370–4379. https://doi.org/10.1103/PhysRevB.6.4370.

(55) Liao, J.; Ji, L.; Zhang, J.; Gao, N.; Li, P.; Huang, K.; Yu, E. T.; Kang, J. Influence of the Substrate to the LSP Coupling Wavelength and Strength. *Nanoscale Res. Lett.* **2018**, *13* (1), 280. https://doi.org/10.1186/s11671-018-2691-2.

(56) Myroshnychenko, V.; Rodríguez-Fernández, J.; Pastoriza-Santos, I.; Funston, A. M.; Novo, C.; Mulvaney, P.; Liz-Marzán, L. M.; Abajo, F. J. G. de. Modelling the Optical Response of Gold Nanoparticles. *Chem. Soc. Rev.* **2008**, *37* (9), 1792–1805. https://doi.org/10.1039/B711486A.

(57) Kluczyk, K.; David, C.; Jacak, J.; Jacak, W. On Modeling of Plasmon-Induced Enhancement of the Efficiency of Solar Cells Modified by Metallic Nano-Particles. *Nanomaterials* **2019**, *9* (1), 3. https://doi.org/10.3390/nano9010003.

(58) Barrow, S. J.; Wei, X.; Baldauf, J. S.; Funston, A. M.; Mulvaney, P. The Surface Plasmon Modes of Self-Assembled Gold Nanocrystals. *Nat. Commun.* **2012**, *3*, 1275. https://doi.org/10.1038/ncomms2289.

(59) McCabe, W.; Smith, J.; Harriott emeritus, P. *Unit Operations of Chemical Engineering*, 7 edition.; McGraw-Hill Education: Boston, 2004.

(60) Kluczyk, K.; Jacak, W. Damping-Induced Size Effect in Surface Plasmon Resonance in Metallic Nano-Particles: Comparison of RPA Microscopic Model with Numerical Finite Element Simulation (COMSOL) and Mie Approach. *J. Quant. Spectrosc. Radiat. Transf.* **2016**, *168*, 78–88. https://doi.org/10.1016/j.jqsrt.2015.08.021.

(61) Amendola, V.; Meneghetti, M. Size Evaluation of Gold Nanoparticles by UV–vis Spectroscopy. *J. Phys. Chem. C* **2009**, *113* (11), 4277–4285. https://doi.org/10.1021/jp8082425.

(62) Fan, J. A.; Wu, C.; Bao, K.; Bao, J.; Bardhan, R.; Halas, N. J.; Manoharan, V. N.; Nordlander, P.; Shvets, G.; Capasso, F. Self-Assembled Plasmonic Nanoparticle Clusters. *Science* **2010**, *328* (5982), 1135–1138. https://doi.org/10.1126/science.1187949.

(63) Dutta-Gupta, S.; Martin, O. J. F. Insight into the Eigenmodes of Plasmonic Nanoclusters Based on the Green's Tensor Method. *JOSA B* **2015**, *32* (2), 194–200. https://doi.org/10.1364/JOSAB.32.000194.

(64) Brandl, D. W.; Mirin, N. A.; Nordlander, P. Plasmon Modes of Nanosphere Trimers and Quadrumers. *J. Phys. Chem. B* **2006**, *110* (25), 12302–12310. https://doi.org/10.1021/jp0613485.

(65) Rechberger, W.; Hohenau, A.; Leitner, A.; Krenn, J. R.; Lamprecht, B.; Aussenegg, F. R. Optical Properties of Two Interacting Gold Nanoparticles. *Opt. Commun.* **2003**, *220* (1), 137–141. https://doi.org/10.1016/S0030-4018(03)01357-9.

(66) Prodan, E.; Radloff, C.; Halas, N. J.; Nordlander, P. A Hybridization Model for the Plasmon Response of Complex Nanostructures. *Science* **2003**, *302* (5644), 419–422. https://doi.org/10.1126/science.1089171.

(67) Halas, N. J.; Lal, S.; Chang, W.-S.; Link, S.; Nordlander, P. Plasmons in Strongly Coupled Metallic Nanostructures. *Chem. Rev.* **2011**, *111* (6), 3913–3961. https://doi.org/10.1021/cr200061k.

(68) Manjavacas, A.; Zundel, L.; Sanders, S. Analysis of the Limits of the Near-Field Produced by Nanoparticle Arrays. *ACS Nano* **2019**. https://doi.org/10.1021/acsnano.9b05031.

(69) Fan, J. A.; Bao, K.; Wu, C.; Bao, J.; Bardhan, R.; Halas, N. J.; Manoharan, V. N.; Shvets, G.; Nordlander, P.; Capasso, F. Fano-like Interference in Self-Assembled Plasmonic Quadrumer Clusters. *Nano Lett.* **2010**, *10* (11), 4680–4685. https://doi.org/10.1021/nl1029732.

(70) Dhiman, M.; Maity, A.; Das, A.; Belgamwar, R.; Chalke, B.; Lee, Y.; Sim, K.; Nam, J.-M.; Polshettiwar, V. Plasmonic Colloidosomes of Black Gold for Solar Energy Harvesting and Hotspots Directed Catalysis for $CO_2$ to Fuel Conversion. *Chem. Sci.* **2019**, *10* (27), 6594–6603. https://doi.org/10.1039/C9SC02369K.

(71) Baffou, G.; Quidant, R.; García de Abajo, F. J. Nanoscale Control of Optical Heating in Complex Plasmonic Systems. *ACS Nano* **2010**, *4* (2), 709–716. https://doi.org/10.1021/nn901144d.

(72) Kim, J. Y.; Kim, H.; Kim, B. H.; Chang, T.; Lim, J.; Jin, H. M.; Mun, J. H.; Choi, Y. J.; Chung, K.; Shin, J.; et al. Highly Tunable Refractive Index Visible-Light Metasurface from Block Copolymer Self-Assembly. *Nat. Commun.* **2016**, *7* (1), 1–9. https://doi.org/10.1038/ncomms12911.

(73) Hamajima, S.; Mitomo, H.; Tani, T.; Matsuo, Y.; Niikura, K.; Naya, M.; Ijiro, K. Nanoscale Uniformity in the Active Tuning of a Plasmonic Array by Polymer Gel Volume Change. *Nanoscale Adv.* **2019**, *1* (5), 1731–1739. https://doi.org/10.1039/C8NA00404H.

(74) Lio, G. E.; Palermo, G.; Caputo, R.; De Luca, A. Opto-Mechanical Control of Flexible Plasmonic Materials. *J. Appl. Phys.* **2019**, *125* (8), 082533. https://doi.org/10.1063/1.5055370.